# Pairing and alpha-like quartet condensation in N=Z nuclei

**N. Sandulescu[1], J. Dukelsky[2] and D. Negrea[1]**

[1]Institute of Physics and Nuclear Engineering, Bucharest-Magurele, Romania
[2]Instituto de Estructura de la Materia, CSIC, Serrano 123, 28006 Madrid, Spain

E-mail: sandulescu@theory.nipne.ro

**Abstract**. We discuss the treatment of isovector pairing by an alpha-like quartet condensate which conserves exactly the particle number, the spin and the isospin. The results show that the quartet condensate describes accurately the isovector pairing correlations in the ground state of systems with an equal number of protons and neutrons.

One of the most debated topic in nuclear physics is the competition between the isoscalar (T=0) and the isovector (T=1) proton-neutron pairing in nuclei with neutrons and protons in the same open shell. The most common formalism to treat both T=1 and T=0 pairing is the generalized Hartree-Fock-Bogoliubov (HFB) approach [1]. The great advantage of this approach is that it can be applied to practically all nuclei, irrespective of their masses and deformations. The drawback is that HFB does not conserve exactly the particle number, the total spin and the total isospin of the nucleus.

A simple alternative to restore the particle number, the spin and the isospin is to work not with pairs but with alpha-like 4-body clusters composed by two neutrons and two protons coupled to total isospin T=0 and total spin J=0 [2]. The existence of alpha-like structures in nuclei, usually related to low-energy alpha emission threshold, regularities in the masses of light nuclei and to specific features of low-energy excitation spectra of N=Z nuclei [3-6], is one of the oldest issue in nuclear structure (for a recent review see [3]).

Various studies rose the question of whether a superfluid condensate of alpha-like quartets could exist in the ground state or in some excited states of self-conjugate nuclei [7-11]. Here some clarifying remarks are in order. In the ground state of nuclei the quartets are not expected to be tight clusters as the alpha particles (nuclei of $^4$He) but rather 4-body structures correlated in angular momentum and isospin space. Thus, in the ground state of heavy nuclei a superfluid condensate of alpha-like quartets is expected to have a small fraction of extended correlated quartets rather than to be a Bose condensate of alpha-like bound quartets. However, according to several calculations, in some specific excited states such as the Hoyle state of $^{12}$C, the quartets could become bound alpha particles and they might do form a boson condensate [12]. A boson condensate of alpha particles has been also predicted in dilute nuclear matter [13].

One of the first microscopic models of quartet condensation in nuclei, proposed by Flowers et al. [7], it is based on a BCS-type function written in terms of quartets. Recently this model was extended by including in the BCS function both Cooper pairs and quartets [11]. A theory of quartet condensation based on a BCS-type function has the advantage of simplicity but its applicability to real nuclei is hindered by the fact that it does not conserve exactly the particle number.

The scope of this paper is to show how the isovector pairing can be described by an alpha-like quartet condensate which conserves exactly the particle number, the spin and the isospin and which is simple enough for performing realistic calculations.

The isovector pairing Hamiltonian is taken in the standard form

$$H = \sum \varepsilon_i (N_i^{(\nu)} + N_i^\pi) - g_1 \sum_{ij,\tau} P_{i\tau}^+ P_{j\tau}, \qquad (1)$$

where the first term is the single-particle Hamiltonian and the second term is the isovector pairing interaction of constant strength. The interaction is expressed by the operators $P_{i\tau}^+ \propto [a_i^+ a_i^+]_\tau^{T=1,J=0}$ where $\tau = \{1,0,-1\} \equiv \{\nu\nu, \pi\pi, \nu\pi\}$ denotes the projections of the isospin corresponding to nn, pp and pn pairs respectively. The Hamiltonian (1) has a local SO(5) symmetry and it can be solved exactly for any distribution of single-particle levels [14]. The most common approximation is the generalized HFB approach [5], which has the drawback that it does not conserve exactly the particle number and the total isospin. An alternative to restore the total isospin is to introduce the quartet operators

$$A_{ij}^+ = [P_i^+ P_j^+]^{T=0} \propto 2 P_{i,\nu\nu}^+ P_{j,\pi\pi}^+ - P_{i,\nu\pi}^+ P_{j,\nu\pi}^+.$$

With these operators we construct a collective quartet operator

$$A^+ = \sum_{ij} x_{ij} (2 P_{i,\nu\nu}^+ P_{j,\pi\pi}^+ - P_{i,\nu\pi}^+ P_{j,\nu\pi}^+). \qquad (2)$$

Because the operator (2) is formed by two neutrons and two protons with the same quantum numbers as in an alpha particle, it will be called an alpha-like quartet. Since the calculations with the collective quartet operator (2) are difficult to perform, we assume that the mixing amplitudes are separable in the indices i and j, i.e., we take $x_{ij} = x_i x_j$. With this ansatz the quartet operator can be written as

$$A^+ = (2 \Gamma_{\nu\nu}^+ \Gamma_{\pi\pi}^+ - \Gamma_{\nu\pi}^+ \Gamma_{\nu\pi}^+), \qquad (3)$$

where $\Gamma_\tau^+ = \sum x_i P_{i,\tau}^+$ are the collective pair operators for the isovector nn, pp and pn pairs.

The quartet operator (3) is used to construct a quartet condensate

$$|\Psi\rangle = A^{+n_q} |0\rangle = \sum_{k=1}^{n_q} (-2)^{n_q-k} \binom{k}{n_q} [\Gamma_{\nu\nu}^+ \Gamma_{\pi\pi}^+]^{n_q-k} [\Gamma_{\nu\pi}^+]^{2k} |0\rangle, \qquad (4)$$

It can be seen that the function (4) is a particular superposition of condensates of collective nn, pp and pn pairs which combine so as to give a total isospin equal to T=0.

The quartet condensate (4) is used as a trial eigenstate for describing the isovector pairing correlations in the ground state of even-even self-conjugate (N=Z) nuclei. Its structure depends on the mixing parameters $x_i$. They are determined by minimizing the average of the Hamiltonian (1) on the quartet condensate (4) and by imposing the normalization condition $\langle\Psi|\Psi\rangle = 1$.

How accurate is the quartet condensate (4) for the description of the ground state of the Hamiltonian (1) can be seen from Table 1 below. Table 1 shows the correlation energies (the differences between the total energies and the HF energies) for a system of 8 proton-neutron pairs distributed in 16 four-fold degenerate single-particle states with the energies $\varepsilon_i = (i-1)/2$. In the third column are shown the results corresponding to the quartet condensate (4) while in the second column are given the exact results [15]. In the last column are shown the results corresponding to the PBCS approximation [15]

$$|PBCS\rangle = (\Gamma_{\nu\nu}^+ \Gamma_{\pi\pi}^+)^{n_q} |0\rangle. \qquad (5)$$

It can be seen that the quartet condensate (4) describes accurately the correlations energies. Thus, for intermediate and strong coupling the errors do not exceed 1%. These errors are comparable with the errors of the PBCS approximation for pairing between like particles [16]. On the other hand, as seen in Table I, the errors corresponding to the quartet condensate are of about one order of magnitude lower than in the case of the PBCS approximation (5) for the isovector pairing Hamiltonian.

Table 1. Correlations energies, in units of the single-particle level spacing, for N=Z=8. In brackets are given the errors relative to the exact results (second column). The third column corresponds to the condensate (4) wile the last to the PBCS approximation (5).

| g | Exact | Quartet | (errors) | PBCS | (errors) |
|---|---|---|---|---|---|
| 0.2 | 1.755 | 1.674 | (4.6%) | 1.239 | (29.4%) |
| 0.4 | 9.917 | 9.820 | (0.9%) | 8.643 | (12.8%) |
| 0.6 | 23.431 | 23.388 | (0.12%) | 21.571 | (7.9%) |
| 0.8 | 39.414 | 39.394 | (0.05%) | 36.968 | (6.2%) |
| 1.0 | 56.586 | 56.574 | (0.02 %) | 53.546 | (5.3%) |

Now we shall discuss the quartet correlations for systems with N=Z=odd. In these systems there is a pn pair which is not included in the quartet condensate. These systems can be described by

$$|\Psi> = \tilde{\Gamma}^+_{\nu\pi} A^{+n_q} |0>, \quad (6)$$

where $\tilde{\Gamma}^+_{\nu\pi} = \sum \tilde{x}_i P^+_{i,\nu\pi}$ describes the residual isovector pn pair. It can be noticed that the odd collective pair has different mixing amplitudes as compared with the pairs which are a part of the quartet condensate. It is expected that the structure of the odd pn pair is quite different to the structure of the pairs belonging to the quartet condensate in the weak coupling limit, and that they tend to be rather similar in the strong coupling regime. In order to examine this feature we show in Table 2 the correlations energies calculated for a system with 7 proton-neutron pairs distributed in 14 double degenerate and equidistant levels. The second column shows the exact results while in the last column are given the results corresponding to the PBCS approximation [15]

$$|PBCS> = (\Gamma^+_{\pi\nu})^7 |0>, \quad (7)$$

which for N=Z=odd gives a lower energy compared to (5). In the third column are shown the results corresponding to the quartet condensate (6) assuming that the odd pair has the same structure as the pairs belonging to the quartet condensate. As seen from Table 2, this assumption is unrealistic for the weak pairing but it works very well in the intermediate and strong coupling regimes.

Table 2. Correlations energies, in units of the single-particle level spacing for N=Z=7. The third column corresponds to the quartet condensate (6); it is supposed that the odd pn pair has the same structure as the pairs in the condensate. The second column gives the exact results while the last corresponds to the results of PBCS approximation (7).

| g | Exact | Quartet | (errors) | PBCS | (errors) |
|---|---|---|---|---|---|
| 0.2 | 1.35 | 0.430 | (68.1%) | 0.585 | (56.7%) |
| 0.4 | 7.59 | 7.496 | (1.2 %) | 3.749 | (54.2%) |
| 0.6 | 17.87 | 17.85 | (0.11%) | 10.73 | (39.9%) |
| 0.8 | 30.99 | 30.99 | (0.002%) | 19.71 | (36.4%) |
| 1.0 | 43.06 | 43.06 | (<0.002%) | 29.65 | (31.1%) |

In conclusion, the present study indicates that an alpha-like quartet condensate describes accurately the isovector pairing correlations in the ground state of the Hamiltonian (1). Preliminary results show that this conclusion is also valid for N=Z nuclei calculated with realistic pairing interactions and single-particle levels [17].

**Acknowledgments** This work has been supported by the Romanian Ministry for Research and Education through the grant PN 09 37 01 02/2010.